\documentstyle[epsfig,aps]{revtex}
\def\beq{\begin{eqnarray}}
\def\eeq{\end{eqnarray}}
\def\bnn{\begin{eqnarray*}}
\def\enn{\end{eqnarray*}}
\def\L{\left(}
\def\R{\right)}

\def\C{{\cal C}}
\def\X{\bar{X}}
\def\n{\bar{n}}
\def\e{\bar{e}}
\def\g{\bar{g}}
\def\d{\delta}
\def\a{\alpha}
\def\aa{a}
\def\M{{\cal M}}

\begin{document}
\title{Fluctuating  brane  in a  dilatonic bulk}
\author{Philippe Brax$^\dagger$, David Langlois$^*$, 
Mar\'{\i}a Rodr\'{\i}guez-Mart\'{\i}nez$^*$}
\address{
${}^\dagger$ Service de Physique Th\'eorique, CEA Saclay, 91191 Gif-sur-Yvette,
France \\
${}^*$ GReCO, Institut d'Astrophysique de Paris (CNRS)
98bis Boulevard Arago, 75014 Paris, France}

\date{\today}
\maketitle

\begin{abstract}
We consider a cosmological brane moving in a static five-dimensional 
bulk spacetime endowed with a scalar field whose potential is exponential. 
After studying various cosmological behaviours for the homogeneous 
background, we investigate the fluctuations of the brane that leave  
spacetime unaffected.  A single mode embodies these fluctuations 
and obeys a  wave equation which we study for bouncing and 
ever-expanding branes. 
\end{abstract}

\section{Introduction}
In the last few years, a lot of efforts have been devoted to  the  
investigation of  the braneworld picture, whereby  
our accessible universe is a three-dimensional submanifold, or three-brane, 
embedded  in a higher-dimensional manifold. 

The cosmological consequences of this idea are of particular interest since 
new effects can be  anticipated  in the very early universe where 
the physical conditions are very different from those of the present universe.
Although many scenarios exist in the literature, most  models   of  brane 
cosmology focus, like  the present work, on 
 a {\it self-gravitating} brane-universe embedded in 
a {\it five-dimensional} bulk spacetime, so that the brane world-sheet 
is of codimension one and subject to the standard junction conditions for 
a thin wall in general relativity. As usually assumed, we will take 
a $Z_2$ symmetric bulk, which means that the two sides of 
the brane are  mirror-symmetric  with 
respect to the brane.

The simplest models of  brane cosmology (see \cite{l02} for a recent review) 
assume an empty bulk with 
a cosmological constant. The latter (with a negative sign) 
is necessary  in 
order to recover a standard cosmological evolution at late times and in 
particular to account, via nucleosynthesis, for the abundances of light 
elements. In the early universe, however, the evolution deviates from 
standard cosmology.

Although very useful for
some specific features of brane cosmology,
an  empty bulk might be too naive for a realistic 
description of the early universe. For example, in the five-dimensional 
version of M-theory, a scalar field, corresponding to the 
volume  of the Calabi-Yau compactification manifold, is present in 
the five-dimensional bulk 
\cite{low}. It is thus relevant  to investigate brane cosmology with 
a bulk scalar field \cite{cr99}-\cite{ftw01}, which might be also useful, 
in the case of two-brane models, to stabilize the radion \cite {gold}. 

In the present work, we consider a five-dimensional model  where the bulk  
contains   a scalar field with an exponential potential and a three-brane 
with a cosmological perfect fluid conformally  coupled, via the bulk 
scalar field, to the induced metric. We restrict our attention to very specific
bulk spacetimes with the usual cosmological symmetries, i.e. homogeneity and 
isotropy along the three ordinary spatial dimensions, that are also 
{\it static}   in the two-dimensional 
subspaces spanned by  time and the extra dimension. We use  {\it static}
here in a generalized sense, where the orbits associated with a
Killing vector  can 
be not only time-like but also space-like. 

We first assume that the brane is perfectly homogeneous and study the 
corresponding background cosmologies  associated with 
 the motion of the brane in such static bulk spacetimes. We thus generalize 
the results of Chamblin and Reall \cite{cr99}, 
restricted to the case of a domain wall, 
to any equation of state for the brane matter. 

We then allow the brane to fluctuate but we impose that 
 these fluctuations are such that 
{\it the bulk spacetime is left unperturbed}. In other words, we investigate
only the fluctuation mode, which one can call the {\it intrinsic mode}, 
that is not coupled to the gravitational radiation, i.e. to the bulk 
perturbations. We show that this mode obeys  a wave equation, which can be 
written in a familiar form. We analyse the evolution of the brane fluctuation
depending on the various background cosmologies.
In some sense, our approach is reminiscent of former studies \cite{gar,guv}
(see also \cite{ishi} for recent developments) of perturbed test branes
where the brane deformation is described by a scalar field obeying a 
Klein-Gordon equation. In our case the self-gravity is included by adjusting
adequately the matter perturbations on the brane.  
A similar
analysis has also been carried out in \cite{steer} within the
context of mirage cosmology where the gravitational back-reaction
is neglected. 
 
The plan of our paper is the following. 
In the next section, we present the framework and consider some background 
homogeneous solutions. In the third section, we derive the 
equation of motion for  the brane fluctuations. In section 4, we analyse 
this wave equation for  the
 background cosmologies discussed in section 2. Finally,  
we conclude in the last section.

\section{The background configuration}

We consider  five-dimensional {\it static} spacetimes with the 
usual cosmological symmetries (homogeneity and isotropy) along the three
ordinary spatial dimensions. The metric can be written in the form
\beq
ds^2 = g_{AB}dx^Adx^B= -A(r)^2 dt^2 + B(r)^2 dr^2 + R(r)^2 d\Sigma^2,
\label{metric}
\eeq
where $d\Sigma^2$ is the metric for maximally symmetric three-dimensional
spaces. For simplicity, we will consider only the flat case.
With our parametrization (\ref{metric}) of the metric, we implicitly assume
that the Killing vector ${\partial\over \partial t}$ is time-like and thus 
that the spacetime is static in the strictest sense. 
Although the calculations below are given explicitely in this context, it is 
not difficult to show that the end results will still hold if the 
coordinate $t$ becomes space-like, i.e. if $A^2$ and $B^2$ 
are negative.

We assume that the bulk contains 
a scalar field $\phi(r)$ with a potential $V(\phi)$.
The five-dimensional action for the {\it bulk} is  given by
an expression of the form 
\beq
{\cal S} = {1 \over  \kappa^2} \int d^5 x \,\sqrt{-g}\,
\left[ {1\over 2} R - \frac{1}{2} \partial^A \phi \partial_A \phi - 
V(\phi)\,\right] \; ,
\label{action}
\eeq
where we have chosen the normalization so that the scalar field is 
dimensionless and the potential scales like a square mass.

The bulk Einstein's equations, derived from this action,  read
\beq
G_{AB} = \partial_A \phi \partial_B \phi 
- g_{AB} \L  {1\over2} \partial^C \phi \partial_C \phi + V(\phi) \R
\eeq
or in terms of the Ricci scalar
\beq
R_{a b} = \partial_a \phi \partial_b \phi + {2\over3} g_{a b}  V(\phi). 
\eeq
Explicitly, they take the form
\beq
\frac{A''}{A} - {A' B' \over A B} + 3 {A' R' \over A R} 
&=& - \frac{2}{3} B^2 V 
\label{einstein1}\\
\frac{R''}{R} + 2 \frac{{R'}^2}{R^2} + \frac{A' R'}{A R} - \frac{B' R'}{B R}
&=& - \frac{2}{3}B^2 V  \\
\frac{A''}{A} + 3 \frac{R''}{R} - {A' B' \over A B} - 3 \frac{B' R'}{B R} &=&
- \frac{2}{3} B^2 V  - {\phi'}^2
\label{einstein3}
\eeq
where a prime denotes a derivative with respect to $r$. 
Similarly the bulk scalar field obeys the Klein-Gordon equation
\begin{equation}
\phi''+(\frac{A'}{A}+3\frac{R'}{R}-\frac{B'}{B^3})\phi'=\frac{\partial V}{\partial \phi}
\label{kg}
\end{equation}
These equations can be solved for specific potentials $V(\phi)$, in particular
for exponential potentials as summarized below.

\subsection{Explicit solutions}
In the case of a scalar field potential of the form
\beq
V(\phi)=V_0 e^{2\alpha \phi},
\eeq
 there exists a simple class of  static solutions\cite{static,cr99}.
The full set of static solutions is given in \cite{charmousis01}, but we 
will restrict our study to the  class of solutions 
 described by  the metric   
\beq
ds^2=-h(R)dT^2+{R^{6\alpha^2}\over h(R)}dR^2+R^2 d{\vec x}^2,
\label{dil_stat}
\eeq
with 
\beq
h(R)=-{ V_0 /6\over 1-(3\alpha^2/4)}R^2-\C R^{3\alpha^2-2},
\eeq
where $\C$ is an arbitrary constant, 
and the  scalar field configuration
\beq
\phi=-3\alpha\ln(R).
\label{phi_stat}
\eeq
Note that, in the limit $\alpha=0$, the scalar field vanishes  while 
its potential reduces to an effective cosmological constant and one recovers 
the well-known Sch-(A)dS five-dimensional metric.
The metric (\ref{dil_stat}) can be expressed in a slightly different
form, as in \cite{cr99}, namely
\beq
ds^2=-U(r)dt^2+{dr^2\over U(r)}+R^2(r)d\Sigma^2,
\eeq
after the change of coordinate
\beq
r=R^{1+3\alpha^2}
\eeq
and a trivial redefinition of time.

\subsection{Moving brane}
Let us now consider the presence of a three-brane moving in the 
static bulk background  (\ref{metric}).
Although we are interested, in this section, only in the motion 
of the homogeneous brane, we already present the general formalism, 
following \cite{ddk00},  
which we will  use later for the study of brane fluctuations.

We define 
the trajectory of the brane in terms of its bulk  coordinates 
$X^A(x^\mu)$ given as
functions of the four parameters  $x^\mu$ which  can be 
interpreted as internal 
coordinates of the brane worldsheet.
One can then define   four independent vectors
\beq
e^A_\mu={\partial X^A\over \partial x^\mu},
\label{e}
\eeq
which are tangent to the brane.
The induced metric on the brane  is simply given by
\beq
h_{\mu\nu}=g_{AB}e^A_\mu e^B_\nu,
\label{h}
\eeq
whereas the extrinsic curvature tensor is given by
\beq
K_{\mu\nu}=e^A_\mu e^B_\nu\nabla_A n_B,
\label{K}
\eeq
where $n^A$ is the unit vector normal to the brane, defined (up to a sign 
ambiguity) by the conditions
\beq
g_{AB}n^An^B=1,\qquad n_A e^A_\mu=0.
\label{n_def}
\eeq
It is also useful to express  $K_{\mu\nu}$ in terms of only 
partial derivatives, which reads
\beq
K_{\mu\nu}={1\over 2}\left[g_{AB}\left(e^A_\mu\partial_\nu n^B+  
e^A_\nu\partial_\mu n^B\right)+
e^A_\mu e^B_\nu n^D\partial_D g_{AB}\right].
\eeq

Let us now  apply this formalism to the homogeneous brane, which can 
be parametrized by
\beq
T=T(\tau), \quad r=r(\tau), \quad X^i=x^i,
\label{param}
\eeq
where we take for the  parameter $\tau$ the proper time, i.e. such that
\beq
h_{\tau\tau}=-1.
\eeq
The induced metric is thus 
\beq
ds^2= -d\tau^2 + R^2(r(\tau)) d\Sigma^2,
\eeq
which shows that the geometry inside the brane is 
FLRW (Friedmann-Lema\^\i tre-Robertson-Walker)  with the scale 
factor given by the radial coordinate $R$ of the brane. The cosmological 
evolution within the brane is thus induced by the motion of the brane 
in the static background.
With the parametrization (\ref{param}), the four independent 
tangent vectors defined in (\ref{e}) take the specific form
\beq
e^A_{\tau} = (\dot T, \dot r, 0,0,0) \;, \;
e^A_{i} = (0,0, \delta_i^A) \;, \;
\eeq
where a dot stands for a derivative with respect to $\tau$
while the components of the normal vector are given by
\beq
n_A= \left(A B \dot r, -B \sqrt{1+  B^2 \dot r^2}, 0, 0,0 \right).
\label{n}
\eeq
Finally, the components of the extrinsic curvature tensor
are  given by
\begin{eqnarray}
K_{ij} &=& - {\sqrt{1 + B^2 \dot r^2} \over B } R R' \, \delta_{ij} \, ,
\label{Kij}
\\
K_{\tau \tau} &=& 
{1 \over A B} {d \over dr} \left( A \sqrt{1 + B^2 \dot r^2} \right)\, .
\label{Ktt}
\end{eqnarray}
Assuming $Z_2$ symmetry about the brane,  the
junction conditions for the metric read 
\beq
K_{\mu\nu}=-{\kappa^2\over 2} \left(S_{\mu\nu}-{1\over 3}S h_{\mu\nu}\right),
\label{junction}
\eeq
where $S_{\mu\nu}$ is the energy-momentum tensor of brane matter and 
$S\equiv S_{\mu\nu}h^{\mu\nu}$ its trace. Because the brane 
is homogeneous and isotropic, $S_{\mu\nu}$ is necessarily of the 
perfect fluid form, i.e. 
\beq
S^\mu_\nu={\rm Diag} \left(-\rho, P,P,P\right),
\eeq
where the energy density $\rho$ and the pressure $P$ are functions of time
only.
Substituting the above  expressions (\ref{Kij}) and (\ref{Ktt}) 
for the components of the 
extrinsic curvature tensor, one finds  
 the  following two relations:
\beq
{\sqrt{1 + B^2 \dot r^2} \over B }{R' \over R} =
{\kappa^2\over 6}\rho,
\label{junction1}
\eeq
and 
\beq
{1 \over A B} {d \over dr} \left( A \sqrt{1 + B^2 \dot r^2} \right) =
- {\kappa^2\over 6}\left(2\rho +3P\right).
\label{junction2}
\end{eqnarray}
Using the first junction condition (\ref{junction1}), the second expression
(\ref{junction2}) 
can be reexpressed explicitly as a conservation-like 
 equation for the energy 
density $\rho$:
\beq
\dot\rho+{\dot R\over R}\left(2 +3 w+f(r) \right) \rho=0,
\label{conservation}
\eeq
where $\dot R\equiv R'\dot r$ and 
\beq
f(r)= {1\over AB}{d\over dr}\left(AB {R\over R'}\right). \label{f}
\eeq
Using Einstein's equations (\ref{einstein1}-\ref{einstein3}), 
this  function $f(r)$ 
can be reexpressed in terms of the scalar field as 
\beq
f(r) = 1 + {1\over 3} {\phi '}^2 \left(R \over R' \right)^2 
\eeq
and the conservation equation takes the form of
\beq
{d\rho\over d\tau}+ 3 {\dot R\over R} \left(\rho + P\right) = 
- {1 \over 3} \rho \dot \phi^2 { R\over \dot R} \, .
\eeq
As we will see in the next subsection, the use of the junction 
condition for the scalar will enable us to  reexpress once more 
this conservation equation in another form.  

\subsection{Junction condition for the scalar field}
In addition to the junction conditions for the metric, we must also ensure 
that the junction condition for the  bulk scalar field is also satisfied.
The latter depends on the specific coupling between $\phi$ and 
the brane matter. In order to be more explicit, we now introduce 
the action for the brane 
\beq
S_{brane}=\int_{brane}d^4x L_m\left[\varphi_m;\tilde h_{\mu\nu}\right],
\eeq
where we assume the metric $\tilde h_{\mu\nu}$ to be 
 conformally related to the induced 
metric $h_{\mu\nu}$, i.e.
\beq
\tilde h_{\mu\nu}=e^{2\xi(\phi)}h_{\mu\nu}.
\eeq
Variation of the total action $S=S_{bulk}+S_{brane}$ 
 with respect to $\phi$ yields the equation 
of motion for the scalar field, 
which is  the Klein-Gordon equation (\ref{kg}) with the addition of    
a distributional source term since  the scalar field is coupled to the brane
via $\tilde h_{\mu\nu}$. An alternative way to deal with this distributional
source term is to reinterpret it as a boundary condition for the scalar 
field, or rather a junction condition at the brane location which takes 
the form 
\beq
\left[n^A\partial_A\phi\right]=\kappa^2 \xi' \left(\rho- 3P\right),
\label{junc_phi}
\eeq
where $\xi'\equiv d\xi/d\phi$. 
Taking into account the $Z_2$ symmetry and the explicit form for the normal
vector (\ref{n}), one ends up with the condition
\beq
\phi'={\kappa^2\over 2} {B \over \sqrt{1 + B ^2\dot r^2}}\xi'(\phi)
\left(-\rho+3P\right),
\label{b_junc_phi}
\eeq
where all terms are evaluated at the brane location.
Moreover, using  the first junction condition (\ref{junction1}), this 
relation can be reduced to  
\beq
\phi'=3\xi'(\phi) {R'\over R}\left(3w-1\right). \label{junction_phi}
\eeq
This junction condition for the scalar field can be substituted in the (non) 
conservation equation (\ref{conservation}) which then reads  
\beq
\dot\rho+3H(\rho+p)=(1-3w)\xi' \rho \dot \phi.
\label{conservation2}
\eeq
This relates the energy loss, from the point of view of the brane, 
to the transverse momentum density, from the point of view of the bulk.  
In fact, this non standard cosmological conservation equation can also 
be rewritten in the standard form 
\beq
{d\tilde\rho\over d\tilde t}+3\tilde H\left(\tilde\rho+\tilde p\right)=0.
\eeq
if one introduces the energy density $\tilde \rho$ and pressure 
$\tilde P=w\tilde \rho$, as well as the scale factor $\tilde a$, 
defined with respect to 
the metric $\tilde h_{\mu\nu}$, which in other contexts would be referred 
to as the {\it Jordan frame}.

\subsection{Brane cosmological evolution}
In order to work with an explicit example, we turn again to the 
dilatonic bulk solutions given in (\ref{dil_stat}-\ref{phi_stat})
 and try to implement a moving 
brane in these backgrounds.

Taking the square of the junction condition (\ref{junction1}), one  
immediately obtains the generalized Friedmann 
equation,
\beq
H^2={\kappa^4\over 36}\rho^2- {h(R)\over R^{2(1+3\alpha^2)}}=
{\hat\rho}^2+{V_0 /6\over 1-(3\alpha^2/4)}R^{-6\alpha^2}
+\C R^{-4-3\alpha^2},
\label{fried}
\eeq 
where we have introduced the notation
\beq
\hat\rho\equiv {\kappa^2\over 6}\rho.
\eeq
For $\alpha=0$, (\ref{fried}) 
reduces to the well-known brane Friedmann equation 
with the characteristic $\rho^2$ term on the right hand side.

As for the scalar field junction condition (\ref{junction_phi}), 
the radial  dependence of 
$\phi$ given in (\ref{phi_stat}) 
imposes the following constraint between the equation of state 
ratio $w$ and $\alpha$:
 \beq
3w-1=-{\alpha\over \xi'}.
\eeq
If we  now assume that $w$ is constant, this  constraint  implies 
that the coupling is linear, i.e. $\xi(\phi)=\xi_1\phi$, in which case the 
conservation equation (\ref{conservation2}) can be explicitly 
integrated to yield
\beq
\rho = \rho_1  R ^{-3(w + 1+\alpha^2)},
\eeq
where $\rho_1$ is a constant. 
One can then substitute this relation into the Friedmann equation 
(\ref{fried})
to obtain 
\beq
\dot R^2+V(R)=0,
\label{energy}
\eeq
with the potential
\beq 
V(R)\equiv -\aa_1 R^{p_1}+\aa_2 R^{p_2}
+\aa_3 R^{p_3}, 
\label{potential}
\eeq
where the coefficients are given explicitly by  
\beq
\aa_1={\hat\rho_1}^2>0, \quad
\aa_2=-{V_0 /6\over 1-(3\alpha^2/4)}, \quad
\aa_3= - \C,
\eeq
and the powers by
\beq
p_1=-4-6w-6\a^2, \quad p_2=2-6\a^2, 
\quad p_3=-2-3\a^2.
\eeq

Equation (\ref{energy}) is analogous to the total energy (which vanishes here)
of a particle moving in a one-dimensional potential $V(R)$. The case of 
a brane domain wall, $w=-1$, was analysed in \cite{cr99}. In this case, 
$p_1=p_2$ and the potential is the sum of only two terms.
Here, however, we have obtained the equation of motion valid  for
 any equation of state of the form $P=w\rho$, with $w$ 
constant. In order to simplify the potential, let us 
 consider the situation $\alpha_3= - \C=0$. It is not difficult, 
from the analysis of the two terms left in the potential $V(R)$, to see 
that the potential has {\it six} distinct shapes, 
 depending  on the sign  
of $\aa_2$ and the value of $\a^2$. To classify  the various cases, 
it is convenient to introduce the parameter $p$ defined, for $w\neq -1$,  by 
\beq
6 \alpha^2 = 2 + p (1 + w) \; .
\label{p}
\eeq
so that the powers $p_1$ and $p_2$ simply read
\beq
p_1=-(6+p)(1+w), \qquad p_2=-p(1+w).
\eeq
The various cases are then:
\begin{itemize}
\item $p<-6$ (which implies $w<-2/3$ and $0<p_1<p_2$) (see fig. 1);
\begin{figure}
\begin{center}
\includegraphics{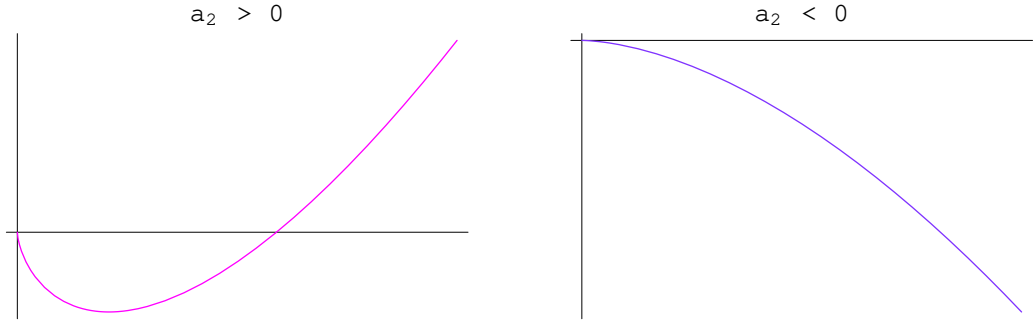} 
\end{center}
\caption{$V(R)$ when $p<-6$. }
\end{figure}
\item $-6<p<0$ (which implies $p_1<0<p_2$) (see fig. 2);

\begin{figure}
\begin{center}
\includegraphics{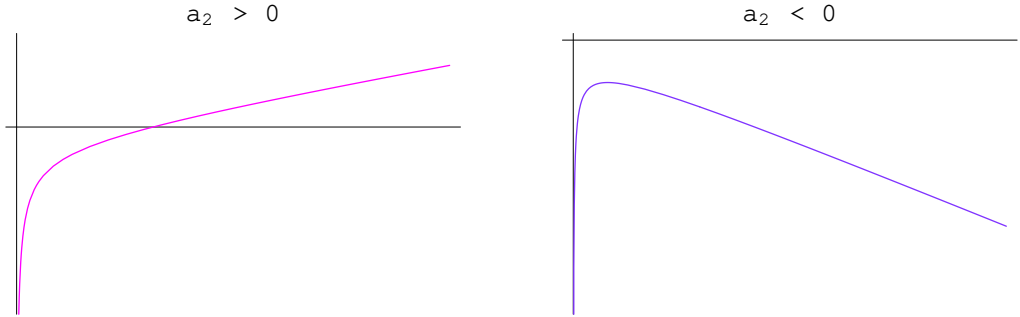} 
\end{center}
\caption{$V(R)$ when $-6<p<0$. }
\end{figure}
\item $p>0$ (which implies $p_1<p_2<0$) (see fig. 3).

\begin{figure}
\begin{center}
\includegraphics{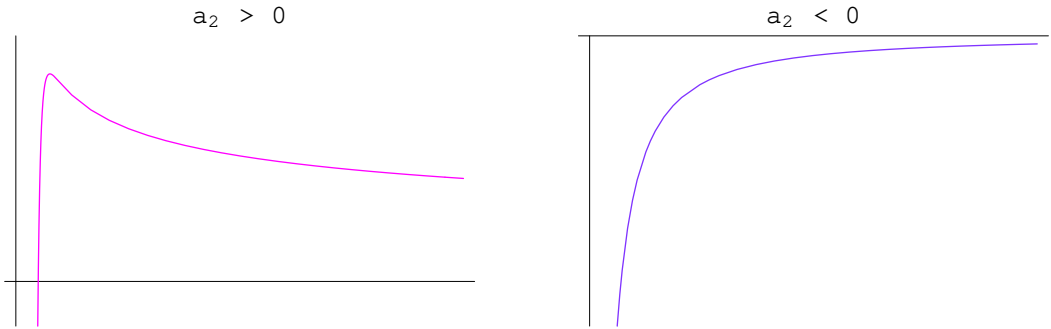} 
\end{center}
\caption{$V(R)$ when $p>0$. }
\end{figure}
\end{itemize}
In the three cases corresponding to $\aa_2<0$, the evolution of the 
scale factor is monotonous because the potential is always negative, 
whereas for $\aa_2>0$, the potential vanishes at a nonzero value $R_c$
which represents the  maximum value of the scale factor during the 
cosmological evolution. In the latter three subcases, cosmological expansion
is  thus followed by a collapse.
This situation $\aa_2>0$ can be seen to be equivalent to the 
supergravity models with a bulk scalar field
and an exponential superpotential \cite{bd}. 
It is also worth noticing 
 that when $\aa_2<0$ the function $h(R)$ parametrizing the metric becomes 
negative. In that case  the coordinate $R$ becomes time-like 
whereas  $t$ becomes space-like and  the Killing vector
corresponding to translations of 
 $t$ is then  space-like. The brane
normal vector 
is then given
 by $n_a= (\sqrt{A^2B^2} \dot r, -\sqrt{B^2+B^4\dot r^2},0,0,0)$ which 
is a real quantity as soon as $\dot r^2$ is large enough. The rest of 
the analysis remains unchanged.

\section{Brane fluctuations}
In this section, we turn to the analysis of the 
{\it brane fluctuations  allowed when the bulk geometry is left unperturbed}.
The fluctuations of the brane will be described by perturbing the 
embedding of the brane in the bulk spacetime, i.e. by writing
\beq
X^A(x^\mu)=\X^A+\zeta \n^A,
\eeq
where the bar stands for the homogeneous quantities defined in the previous 
section.
The four tangent vectors defined in (\ref{e}) are then given by
\beq
e^A_\mu=\bar{e}^A_\mu+\d e^A_\mu=\bar{e}^A_\mu+
\partial_\mu\left(\zeta \n^A\right).
\label{e_p}
\eeq
Substituting in the definition of the induced metric (\ref{h}), 
and being careful  to 
evaluate  the ({\it unchanged}) bulk metric $g_{AB}$ at the {\it perturbed}
 brane location, one finds 
\beq
h_{\mu\nu}=\bar{h}_{\mu\nu}+2\zeta \bar{K}_{\mu\nu}.
\label{h_p}
\eeq
Using this expression, one can  easily make the connection
with  the Bardeen potentials measuring the gauge invariant
metric perturbations induced by the fluctuations of the brane position.
In the  longitudinal gauge, the perturbed metric reads 
\begin{equation}
ds^2= -(1-2\Psi) dt^2 + R^2(1+2\Phi) \delta_{ij}dx^idx^j,
\end{equation}
and by comparing with (\ref{h_p}), one finds  that
\begin{equation}
\Psi=-\zeta \bar K^{\tau}_{\tau},\qquad \Phi={1\over 3}\zeta \bar K^i_i
\end{equation}
which gives, after using the background junction conditions (\ref{junction}),
\begin{equation}
\Psi=-\frac{\kappa^2}{6}(2+3w)\rho\zeta
\label{bardeen_pot_1}
\end{equation}
and
\begin{equation}
\Phi=-\frac{\kappa^2}{6}\rho\zeta
\label{bardeen_pot_2}
\end{equation}
The metric perturbations are thus directly proportional to   the brane
fluctuation $\zeta$. We will return later  to the evolution of 
 the Bardeen potentials. The rest of this section is devoted to the 
derivation of the equation of motion that governs the evolution of the 
brane fluctuation. We first consider the perturbed junction conditions 
for  the metric and then those for the scalar field. 

\subsection{Perturbed junction conditions for the metric}

As a first step, let us evaluate the perturbed normal vector, which can 
always be decomposed as
\beq
\d n^A=\alpha \n^A+\beta^\mu \e^A_\mu.
\label{n_p}
\eeq
The coefficients $\alpha$ and $\beta^\mu$ can be determined by perturbing
the two equations in (\ref{n_def}). They are given by
\beq
\alpha=-{1\over 2}\zeta \n^A\n^B\n^C \partial_C\g_{AB}
\eeq
and 
\beq
{\bar h}_{\mu\nu}\beta^\nu=-\zeta \n^A \e^B_\mu\n^C\partial_C \g_{AB}
-\g_{AB}\n^A \partial_\mu\left(\zeta \n^B\right).
\eeq
Substituting the expressions (\ref{e_p}) and (\ref{n_p}) in the perturbation
of the extrinsic curvature tensor (\ref{K}), one obtains  the expression
\begin{eqnarray}
\d K_{\mu\nu}&=&{1\over 2}\left[ 2
\zeta\n^C\partial_C \g_{AB}\e^A_{(\mu} \e^B_{\nu)}
+2\g_{AB}\left(\d e^A_{(\mu}\partial_{\nu)} \n^B+  
\e^A_{(\mu}\partial_{\nu)} \d n^B\right) +
\right.\cr
&& \left.
+\left(\d e^A_\mu \e^B_\nu \n^C+
\e^A_\mu \d e^B_\nu \n^C+ \e^A_\mu \e^B_\nu \d n^C
\right)\partial_C \g_{AB}
\right.\cr
&& \left.
+\zeta
\e^A_\mu \e^B_\nu \n^C\n^D\partial_C\partial_D \g_{AB}\right].
\end{eqnarray}
The expression with an upper index and a lower index is also useful and 
can be obtained from the above expression by using the relation
\beq
\d K_\mu^\nu=\bar{h}^{\nu\sigma}\d K_{\sigma\mu}- 2\zeta \bar{K}_{\mu\sigma}
\bar{K}^{\sigma\nu},
\eeq
where the indices for $\bar{K}^{\sigma\nu}$ are raised by using the 
inverse metric $\bar{h}^{\rho\sigma}$.

The explicit evaluation of the components of the perturbed brane 
extrinsic curvature
tensor,  for the metric (\ref{metric}), then  yields
\begin{eqnarray}
\d K^\tau_\tau &=& 
\ddot \zeta -\left[ 
{\L B \ddot r + B' \dot r^2 \R^2 \over 1 + B^2 \dot r^2} +
 \L  1 + B^2 \dot r^2 \R {{A'}^2 \over A^2 B^2}
+{A' B' \over A B^3}
\right. \cr \label{dKtt}
&&
\left.
+2   {A' B' \over A B} \dot r^2 - {A'' \over A B^2} +
2  {A' \over A} \ddot r\right] \zeta,
\\
\d K^\tau_i &=&  \dot \zeta_{,i} -  { R' \over R}\dot r \zeta_{,i}, 
\label{dKti} \\
\d K^i_j &=&\left[  {R' \over R}  \dot r \dot \zeta
+\left(  
-  R' B' {1 + B^2 \dot r^2\over B^3 R} 
-  \dot r^2 {A' R' \over A R} 
\right. \right. \cr 
&&\left.\left.
- {R'} ^2 {1 + B^2 \dot r^2 \over B^2 R^2} + 
 R''{ 1 + B^2 \dot r^2 \over B^2 R} \right) \zeta\right]\delta^i_j 
-{\zeta_{,j}^{,i} \over R^2} 
\label{dKij}
\end{eqnarray}
In the longitudinal gauge, which we shall use, the components of the
perturbed brane energy momentum tensor read
\begin{eqnarray}
\d S^\tau_\tau&=&-\d\rho, \label{dS1}\\
\d S^\tau_i &=& R(1+w)\rho\partial_i v,\\
\d S^i_j &=& \d P\d^i_j+\d\pi^i_j, \label{dS3}
\end{eqnarray}
where 
\beq
\d\pi^i_j=\d\pi_{,j}^{,i}-{1\over 3}\delta^i_j \d\pi_{,k}^{,k}
\eeq
is the (traceless) anisotropic stress tensor, 
and the perturbed junction conditions for the metric, which follow 
from (\ref{junction}), are given by 
\beq
\d K^\mu_\nu=-{\kappa^2\over 2}\left(\d S^\mu_\nu-{\d S\over 3}\d^\mu_\nu\right).
\eeq
Inserting (\ref{dS1}-\ref{dS3}),  this gives explicitly
\begin{eqnarray}
\d K^\tau_\tau&=&{\kappa^2\over 6}\left(2\d\rho+3\d P\right),
\label{d_junction1}
\\
\d K^\tau_i&=&-{\kappa^2\over 2}R(1+w)\rho\partial_i v,
\label{d_junction2}
\\
\d K^i_j &=&-{\kappa^2\over 2}\left( {1\over 3}\d \rho\d^i_j+\d\pi^i_j\right).
\label{d_junction3}
\end{eqnarray}
The second equation (\ref{dKti}) determines, once $\zeta$ is known, the velocity potential 
$v$, except when the equation of state is $w=-1$, in which case one gets 
the constraint
\beq
\dot \zeta= {R'\over R}\dot r \zeta \qquad (w=-1).
\eeq
This implies that the perturbation reads
\begin{equation}
\zeta= R(\tau) C(k)
\end{equation}
up to a global translation of the brane. The function $C(k)$ will be determined later.

Finally,
equation (\ref{d_junction3}) can be 
decomposed into a  trace and a traceless part, giving respectively 
\begin{eqnarray}
 {R' \over R}  \dot r \dot \zeta
+\left(
-  R' B' {1 + B^2 \dot r^2\over B^3 R} 
-  \dot r^2 {A' R' \over A R} - {R'} ^2 {1 + B^2 \dot r^2 \over B^2 R^2}
\right.
\nonumber \\
\left.
 + 
 R''{ 1 + B^2 \dot r^2 \over B^2 R} \right) \zeta
 - {1 \over 3 R^2} \Delta \zeta
= -{\kappa^2 \over 6}\delta \rho 
\label{eq2_zeta} 
\end{eqnarray}
and 
\begin{equation}
{1\over R^2}
\left(\zeta_{,j}^{,i}-{1\over 3}\Delta \zeta \delta^i_j \right) = {\kappa^2 
\over 2}\delta \pi^i_j. 
\end{equation} 
The last equation simply gives
\begin{equation}
\delta\pi= \frac{2}{\kappa^2}\frac{\zeta}{R^2}
\label{pi}
\end{equation}
and shows  that the anisotropic stress is intrinsically related to the 
brane fluctuation.

\subsection{Perturbed junction condition for the scalar field}
The next step in order to establish the equations of motion for the brane 
fluctuations is to write down the perturbed 
junction condition for the scalar field.
The first order perturbation of 
(\ref{junc_phi}) yields 
\beq
\left[\delta n^A \partial_A \phi  + 
\zeta n^A n^B \partial_A \partial_B \phi \right] 
= - \kappa^2 \,\xi'\, \delta S - \kappa^2 S \,\xi'' 
(n^A \partial_A \phi)\zeta. 
\label{jonction_phi_pert}
\eeq
Taking into account $Z_2$ symmetry and using the background junction
condition (\ref{b_junc_phi}), its derivative along the trajectory, and  
the other junction condition (\ref{junction1}), one finds, after some 
algebra, that   eq. (\ref{jonction_phi_pert}) takes the form 
\begin{eqnarray}
&&\left(3w-1\right)\left\{{R'\over R}\dot r \dot\zeta+
\left[{\hat\rho}^2\left({RR''\over {R'}^2}-1\right)
-{R'\over R}\left({B'\over B^3}+\left({A'\over A}+{B'\over B}\right)\dot r^2\right)\right]\zeta\right\}\cr
&&
+3{\hat\rho}^2{R\over R'}{\dot w\over \dot r}\zeta=
{\kappa^2\over 6}\left(1-3c_p^2\right)\delta\rho
\label{eq1_zeta}
\end{eqnarray}
where we have introduced
\begin{equation}
c_p^2=\frac{\delta p}{\delta \rho}
\end{equation}
Combining (\ref{eq1_zeta}),  (\ref{eq2_zeta}) and 
(\ref{d_junction1}), one sees that the matter perturbation can 
be eliminated to give a differential equation that depends only on
$\zeta$. It has the form of
 a wave equation. 
and  reads 
\beq
\ddot \zeta + (2 + 3w){\dot R \over R} \dot\zeta  
- {\Delta \zeta \over R^2}+\left\{
 {A'' \over A B^2} - {A' B' \over A B^3}- 
(2 + 3w){A' R' \over A R}  \dot r^2 \right.
\nonumber
\\
\left.
+(2+3w){\hat\rho}^2\left[-(2+3w)-{B' R' \over B R}+{R R''\over {R'}^2}-1
\right]\right\}\zeta=0.
\label{ec_ondas}
\eeq

Introducing the function $\psi$ defined by
\beq
\psi=R^{(1+3w)/2}\zeta
\eeq
and using the conformal time $\eta$ defined by  $d\tau= Rd\eta$, one can 
rewrite  the wave equation in the simple form 
\begin{equation}
{d^2\psi\over d\eta^2} +(k^2+{\cal M}^2) \psi=0,
\label{wave}
\end{equation}
where the effective mass is given by 
\begin{eqnarray}
{\cal M}^2&=&
R^2 \left[-\frac{1+3w}{2}\left(\frac{\ddot R}{R}+\frac{1+3w}{2}\frac{\dot
R^2}{R^2}\right)
+\frac{A''}{AB^2}-\frac{A'B'}{AB^3}-(2+3w)\frac{A'R'}{AR}\dot r^2  
\right.
\nonumber \\
&& \left. + (2+3w){\hat\rho}^2
 \left(\frac{RR''}{R'^2}-\frac{B'R}{BR'}-3(1+w)\right)\right].
\label{m2_gen}
\end{eqnarray}

We have thus obtained the wave equation governing the intrinsic brane 
fluctuations in the general case. Initially, we started from a system of 
five equations, (\ref{d_junction1}), (\ref{d_junction2}), 
(\ref{eq1_zeta}), (\ref{pi}) and  (\ref{eq2_zeta}), all obtained from the 
junction conditions, either of the metric or of the scalar field. 
These five equations contain one dynamical equation, 
which has been expressed above in terms of the quantity 
$\zeta$ (or $\psi$) and four constraints which yield respectively 
the energy density $\delta\rho$, the pressure $\d P$, the 
four-velocity potential $v$ and the anisotropic stress $\d\pi$. In contrast
with the standard cosmological context where one can choose beforehand 
 the relation between  $\d P$ and $\d\rho$, and the anisotropic stress, 
they are here completely determined by the constraints  once a solution 
for $\zeta$ is given. This is necessary to get  
a configuration where the brane is fluctuating while the background 
is unaffected. Intuitively, this means that 
the gravitational effect  due to the geometrical 
fluctuations of the brane must be exactly compensated 
 the distribution of matter in the brane, so that the 
net  gravitational effect due to the presence of 
  the brane is completely cancelled in the bulk.  

In the rest of the paper, we will specialize
our study to specific solutions, which will simplify the expression of
the effective mass. 

\section{Perturbations in Dilatonic Backgrounds}
In this section we will focus on the dilatonic backgrounds
described earlier, corresponding to exact solutions for an exponential 
potential. 
Using the previous general result about the mass term ${\cal
M}^2$ in the wave equation, we can now specialize these results to
the dilatonic backgrounds.  We will concentrate on the
case where the background equation of state parameter $w$ is constant.
Substituting the solution (\ref{dil_stat}) in the expression 
(\ref{m2_gen}), one finds 
that the square mass reads
\begin{eqnarray}
{\cal M}^2&=&\frac{R^2}{4}
\left\{(1+w)\left[
{2V_0\over 3\alpha^2-4} (5+3w-6\alpha^2)R^{-6\alpha^2}-3
(7+9w+6\alpha^2)\hat{\rho}^2\right]
\right.\cr
&&\left. 
+3(1-w)(3\alpha^2 -1+3w) {\cal C}R^{-4-3\alpha^2}\right\}
\end{eqnarray}
where
\begin{equation}
\hat \rho^2=\frac{1+B^2\dot r^2}{B^2}(\frac{R'}{R})^2
\end{equation}
Notice that for $w=-1$ only the ${\cal C}$ dependent term remains.
Using the decomposition $\zeta=C(k) R$ we find that
\begin{equation}
(k^2+{\cal M}^2)C(k)=0
\end{equation}
which leads to $C(k)=0$ and therefore the absence of brane
fluctuations for $w=-1$. 

In the following we will concentrate on the ${\cal C}=0$ case. 
Introducing the parameter $p$ defined in (\ref{p}), the squared mass 
is now given by 
\begin{eqnarray}
{\cal M} ^2 &=& 
{3\over 4}(1 + w)^2\left[-\aa_1(9+p)R^{p_1}+\aa_2 (3-p) R^{p_2}\right]\cr
&=&
(1 + w)^2 \frac{V_0}{2(3\alpha^2-4)}\left[ {(3 - p) \over 
R ^ {p(1 + w)}}
 -\delta { (9 + p) \over R ^ {(6 + p )(1 + w)} } \right], 
\end{eqnarray}
with 
\begin{equation}
\delta= \frac{3(3\alpha^2-4)}{2V_0} \hat\rho_1^2.
\end{equation}
We will now treat separately the   cases  of positive or negative 
$\aa_2$, i.e. of $\delta$, which correspond to very different behaviours. 

\subsection{Bouncing branes}
Let us concentrate first on the case
\begin{equation}
\aa_2\equiv\frac{2V_0/3}{3\alpha^2-4}=\gamma^2>0,
\end{equation}
where $\gamma$ has the dimension of  mass. 

The motion can be conveniently analysed by defining 
 $y = R^{6(w+1)}$.
The equation of motion (\ref{energy}) yields
\beq
{dy \over 6(1 + w) \sqrt{\delta - y}} y^{-\frac{1}{2}+\frac{p}{12}}
= \pm \gamma  d\eta 
\eeq
Let us define $n $ as 
\beq
n = \frac{p-6}{12}
\eeq
The scale factor is then given by the implicit relation
\beq
{y^{1+n} \over 1+n } {\mathcal{F}}(1+n,{1 \over 2},2+n,{y \over \delta}) = 
\pm 6 \gamma(1 + w)  \sqrt{\delta} \, \eta + \eta_0
\eeq
${\mathcal{F}}$ being the hypergeometric function.
Of course the scale factor is only determined after inverting
these equations. 
The motion is bounded from above by $y=\delta$.
For a brane whose scale factor increases initially, it reaches a
maximal value corresponding to $\delta$ before bouncing back and being
irredeemingly attracted by the singularity located at $R=0$.
It is interesting to notice that for $p\le -6$ the singularity 
is reached at infinite conformal time while for $p>-6$ it takes
a finite amount of conformal  time to the brane in order to reach 
the singularity.

Let us now analyse the square mass driving the brane fluctuations.
We have plotted the different cases in Figure 4. There is a qualitative
change of behaviour for the square mass when $p>-6$.
Below the critical value $p=-6$ the mass vanishes at the
singularity when $R$ vanishes. This leads to an oscillatory
behaviour of the brane fluctuation. Above that threshold the
squared mass becomes infinitely negative at the singularity
leading to an instability of the brane to fluctuations, i.e. the brane
tends to be ripped to shreds by the presence of the singularity.
\begin{figure}
\begin{center}
\vspace*{-1.5cm}
\epsfxsize=3in
\epsfysize=3in
\begin{picture}(450,200)
\includegraphics{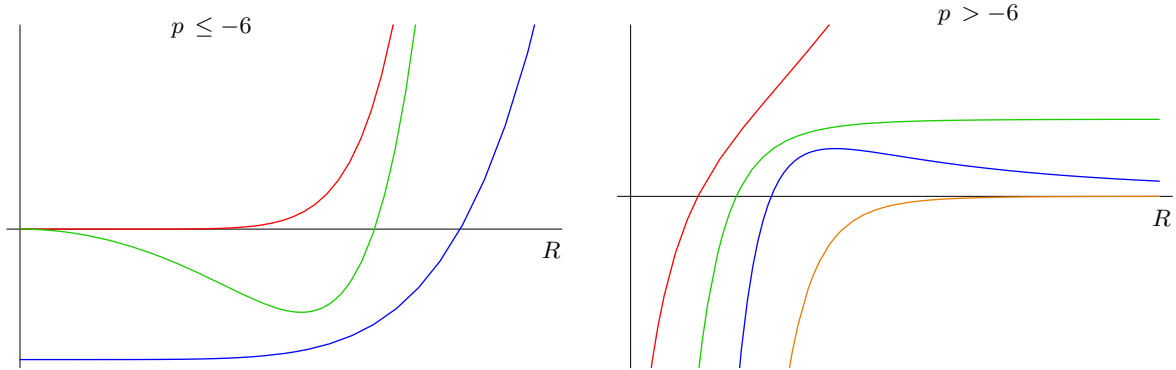}
\put(-390,130){\small{$p \, \leq -6 $}}
\put(-100,135){\small{$p \, > -6 $}}
\put(-250,45){\small{$R$}}
\put(-19,57){\small{$R$}}
\end{picture}
\end{center}
\vspace*{0.8cm}
\caption{ ${\mathcal{M}}^2$ in a bouncing universe. In the left picture 
($p \le -6$), the red line corresponds to $p \le -9$, the green line 
represents  $-9 <p < -6$ and the
blue line displays the critical value $p = -6$. 
The right picture shows 
${\mathcal{M}}^2$ for  $p > -6$ : the red line corresponds to $-6 <p < 0$,
the green line to $p = 0$, the blue line stands for $0 <p < 3$ whereas 
the yellow line stands for $p \ge 3$.}
\label{mebb}
\end{figure}

In the cases $-9<p<-6$, the square  mass is negative for small values of 
$R$, reaches a minimum and then increases up to positive values 
with increasing $R$. However, values of $R$ greater than 
$\delta^{1/(6(w+1))}$  are irrelevant since the background evolution 
bounces when  reaching this maximum value. The position of the minimum is 
given by 
\beq
y_m={(6+p)(9+p)\over p(3-p)}\delta <\delta,
\eeq
whereas  the scale factor corresponding to $\M^2=0$ is given by
\beq
y_0={(9+p)\over (3-p)}\delta.
\eeq
For $-6<p<3$, the square mass starts from negative values and becomes 
positive after the critical value $y_0$
which is less than $\delta$ only for $p<-3$. In other words, for cases 
$p>-3$, the region corresponding to positive square mass is irrelevant.

We can recover this qualitative analysis by studying the
solutions of the wave equation (\ref{wave}), which, in terms of the variable
$y$, reads 
\beq
{d^2 \psi \over dy^2} \L \delta -y \R + {d \psi \over dy } 
\L n -{1 \over 2} -{ n \delta \over y} \R + 
y^{2n} \tilde k^2 \psi 
= {y \L4 n + 1\R + \delta \L4 n+5\R  \over{16 y^2}} \psi
\eeq
with 
\beq
 \tilde k^2 = {k^2 \over 36 \gamma^2 \L 1 + w \R^2 }
\eeq
The variable $y$ evolves between $0 < y < \delta$.

The asymptotical behaviour of the perturbation near the
singularity $y=0$ depends on $n$ :
\beq
\psi \sim \left\{
\begin{array}{ll}
{\mathcal{C}}_1 y^{-{1\over4}} + {\mathcal{C}}_2 y^{n + {5\over4}}, & 
 \; n > -1\\
{\mathcal{C}}_1 \cos 
\L \sqrt{{ \frac{\tilde k^2}{\delta}  - \frac{1}{16} }} \ln \, y \R + 
{\mathcal{C}}_2\sin 
\L \sqrt{{ \frac{\tilde k^2}{\delta} - \frac{1}{16} }} \ln \, y \R,  
&  \; n = -1 \\
{\mathcal{C}}_1 \cos  \L y^{n+1} +\alpha  \R +
{\mathcal{C}}_2 \sin  \L y^{n+1} +\alpha  \R, &  \; n < -1
\end{array} \right.
\label{comp_asint}
\eeq

Notice that for  $p<-6$  the
brane oscillates for an infinite amount of conformal time
before reaching the singularity.
For $p>-6$, the brane stops oscillating and hits the singularity
in a finite amount of conformal time. For $p= -6$
the brane oscillates only for small length-scales
corresponding to $\tilde k> 4 \sqrt \delta$.

The Bardeen potentials, given in (\ref{bardeen_pot_1}) and 
(\ref{bardeen_pot_2}), are also worth investigating. They are proportional,
\beq
\Psi = (2+3w) \Phi,
\eeq
and related to the brane fluctuation according to 
\beq
\Phi \propto R^{{(9+p) \over 2 (1+w)}} \psi.  
\eeq
One thus notices the critical value $p=-9$, above  which 
the Bardeen potentials are enhanced, for an expanding universe, with respect 
to $\psi$. 
One can compute numerically the evolution of the perturbation, 
the Bardeen potential and the scale factor
as a function of the conformal time (see Figure 5).

\begin{figure}
\begin{center}
\vspace*{-1.5cm}
\begin{picture}(470,200)
\includegraphics{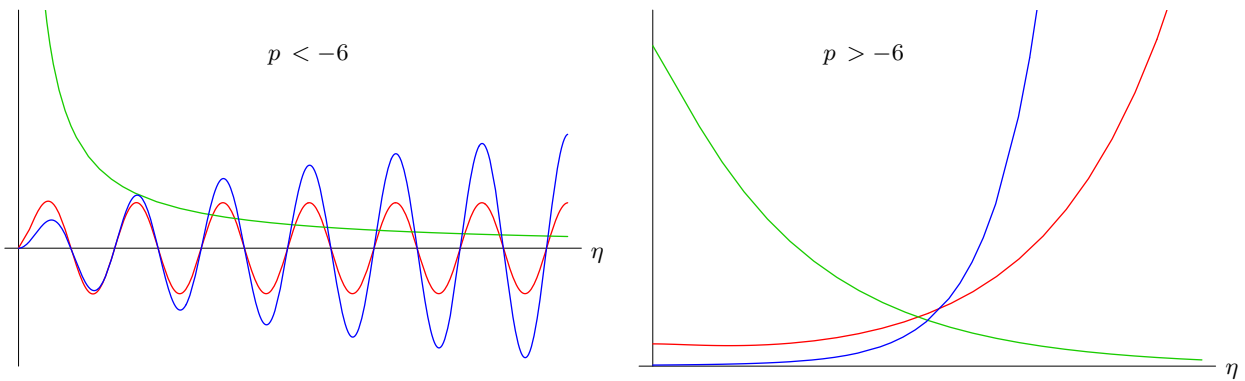} 
\put(-370,120){\small{$p \, < -6 $}}
\put(-160,120){\small{$p \, > -6 $}}
\put(-248,45){\small{$\eta$}}
\put(-8,1){\small{$\eta$}}
\end{picture}
\vspace*{0.8cm}
\caption{Perturbation (red line), Bardeen potential (blue line) and scale 
factor (green line)  during the contracting phase of a bouncing universe, 
as functions of the conformal time.}
\label{bb}
\end{center}
\end{figure}

\subsection{Ever expanding  branes}
We now turn to the case
\begin{equation}
\aa_2\equiv\frac{2V_0/3}{3\alpha^2-4}=-\gamma^2<0,
\end{equation}
Using once more the $y$ variable we can rewrite the 
background evolution equation
as
\beq
{dy \over 6(1 + w) \sqrt{y+|\delta|}} y^{-\frac{1}{2}+\frac{p}{12}}
= \pm \gamma  d\eta. 
\eeq
\begin{figure}
\begin{center}
\vspace*{1.5cm}
\begin{picture}(470,200)
\includegraphics{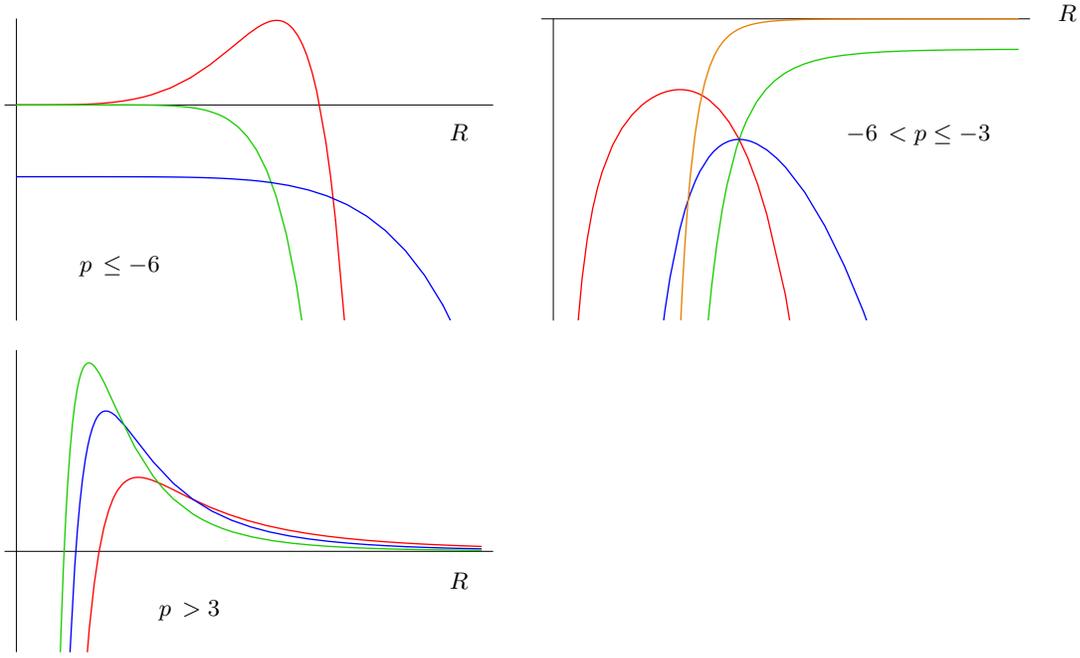} 
\put(-370,150){\small{$p \, \leq -6 $}}
\put(-80,200){\small{$-6 \, < p \leq  -3 $}}
\put(-340,20){\small{$p \, >3 $}}
\put(-230,200){\small{$R$}}
\put(0,245){\small{$R$}}
\put(-230,30){\small{$R$}}
\end{picture}
\vspace*{0.8cm}
\caption{${\mathcal{M}}^2$ in an inflationary universe. 
The picture $p \le -6$
depicts three different cases : $p < -9$ (red line),  $-9 \le p < -6$ 
(green line) and $p = -6$ (blue line).
In the second picture, where $-6 < p \le -3$, the red and blue lines 
illustrate $-6 < p < 0$, the green line stands for $p = 0$ 
and the yellow line stands for $0 <p \le 3$. The third picture represents
$p > 3$.}
\end{center}
\end{figure}
Since $p_2>p_1$,  the asymptotic behaviour at early times, i.e. at small $R$, 
is dominated, both for the background and for the perturbations, 
 by the $R^{p_1}$ term 
(since $p_2>p_1$), which does not depend on the sign of $\aa_2$. Therefore, 
the asymptotic behaviour at early times for ever expanding branes is 
exactly the same as that found in the case of bouncing branes. 

For large $R$, conversely, the dominating term is $R^{p_2}$. 
For  the background, this leads to a power-law behaviour of the scale 
factor, explicitly given by
\beq
R(\eta)\simeq\left[{p\over 2}(1+w)\gamma \eta\right]^{2\over p(1+w)},
\label{power-law}
\eeq
which, 
in terms of the cosmic time,  translates into
\beq
R(\tau)\propto \tau^{2/\left(p(1+w) +2\right)}.
\eeq
As soon as $p<0$, one gets an accelerated expansion, similar to the 
standard four-dimensional 
power-law inflation, which can be obtained from a scalar field 
with an exponential potential. For $p>0$, one gets a decelerated power-law
expansion. 
It is instructive to compare the power-law expansion for the brane
with the standard expansion law, which is given 
by
\beq
R(\tau)\propto \tau^{2/3(1+w)}.
\eeq
Substituting the expression  (\ref{power-law}) in the squared mass, one 
finds 
\beq
\M^2=-{3(3-p)\over p^2}{1\over\eta^2}.
\eeq 

Note that, in the case of power-law inflation, one can derive 
a second-order differential equation of the form (\ref{wave}) 
for a canonical variable  which is a linear combination of the scalar 
field perturbation and of the (scalar) metric perturbation. For a power-law
$a\sim t^q$, one would find 
\beq
\M^2_{eff}=-{q(2q-1)\over (q-1)^2}{1\over \eta^2}.
\eeq
It is easy to check that our expression for $\M^2$ does not coincide 
with the $\M^2_{eff}$ deduced from power-law inflation, for the same 
evolution of the background. 
With power-law inflation, the spectrum for the Bardeen potential(s) is 
given by
\beq
{\cal P}\sim k^{-2/(q-1)},
\eeq  
which tends to a scale-invariant spectrum for large power $q$. 

In our case,  we obtain that the fluctuations are
\begin{equation}
\psi=\sqrt{-\eta}\left[\a_1 H^{(1)}_\nu (-k\eta) + \a_2 H^{(2)}_\nu (-k\eta)
\right]
\end{equation}
where
\begin{equation}
\nu= \frac{p-6}{2p}
\end{equation}

If one assumes that  $\psi$ is given in the asymptotic past $\eta\rightarrow 
-\infty$ as  the usual vacuum solution in inflation, i.e.
\beq
\psi\sim {1\over \sqrt k}e^{-ik\eta}, \qquad k|\eta|\gg 1,
\eeq
then this means that $\a_2=0$ and the behaviour on long wavelengths is given
by 
\beq
\psi\sim (-\eta)^{{1\over 2}-\nu}k^{-\nu}\sim R^{{3\over 2}(1+w)}k^{-\nu},
\qquad k|\eta|\ll 1.
\eeq
  Using the relation between $\psi$ and the Bardeen potential, one thus finds
that the spectrum for $\Phi$ is given by
\beq
{\cal P}_\Phi\sim R^{9+p+3(1+w)^2\over 1+w}k^{-2\nu}.
\eeq
Contrary to the four dimensional inflationary case, the spectrum of the 
Bardeen potential
is not constant outside the horizon. Moreover the  spectrum is red and far 
from being scale-invariant.
Hence, despite an inflationary phase on the brane, the intrinsic 
fluctuations of a brane in a dilatonic background
are not a candidate for the generation of primordial fluctuations. 
\begin{figure}
\begin{center}
\vspace*{-1.5 cm}
\begin{picture}(470,200)
\includegraphics{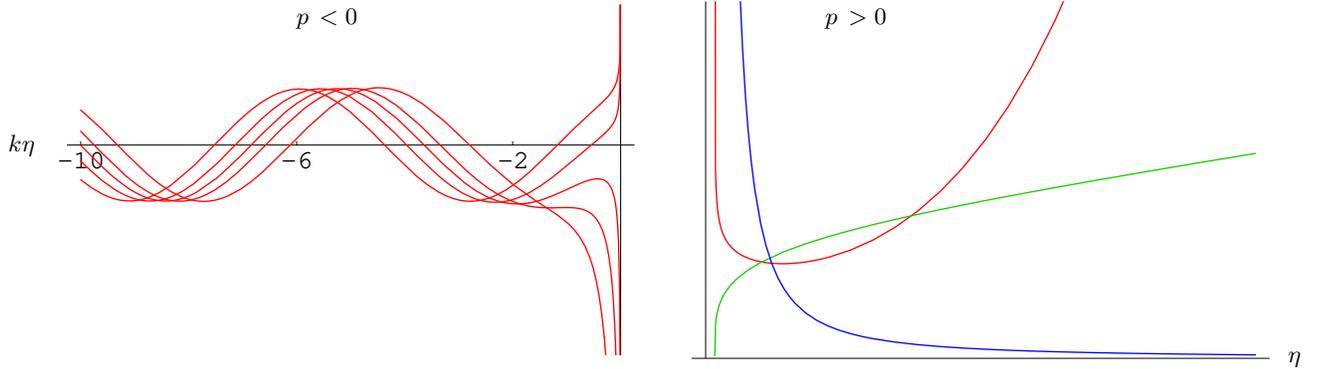} 
\put(-380,130){\small{$p \, < 0 $}}
\put(-180,130){\small{$p \, > 0 $}}
\put(-489,82){\small{$k \eta$}}
\put(-5,2){\small{$\eta$}}
\end{picture}
\vspace*{0.8cm}
\caption{Left picture : Perturbation as a function of $k \eta$. 
For $p < 0$, the perturbation
always diverges near the singularity. Right picture : perturbation 
(red line), Bardeen potential (blue line) and scale 
factor (green line ) for $p > 0$.}
\end{center}
\end{figure}

\section{Conclusion}

We have investigated the fluctuations of a moving brane in a
dilatonic background. These fluctuations are represented by a
scalar mode on the brane corresponding to ripples 
along the normal direction to the brane. As the brane fluctuates,
it induces metric fluctuations, in particular we have found that
the induced metric appears naturally in the longitudinal gauge
with two unequal Bardeen potentials $\Phi$ and $\Psi$. The fact
that these potentials are not equal springs from the presence of
anisotropic stress on the brane. For a fixed equation of state
for the matter content on the brane, for instance cold dark
matter, we find that the two Bardeen potentials are proportional.
As such this implies that a single gauge invariant observable
$\Phi$ 
characterizes the brane fluctuations.

 Our approach differs from the projective approach \cite{mw00,mb00} 
in as much as
we have not considered the perturbed Einstein equations on the brane.
This allows us to free ourselves from the thorny problem of the
projected Weyl
tensor on the brane.

We have focused on the motion and fluctuations of branes in a
particular class of dilatonic backgrounds. These backgrounds
correspond to an exponential potential and an exponential coupling
of the bulk scalar  field to the brane. The motion of the brane
is either of the bouncing type  or  the ever-expanding form.
In the bouncing case we find that the brane cannot escape towards
infinity, it is bound to a singularity which is either null or time-like.
In the time-like case, i.e. when it appears at a  finite distance in
conformal coordinates, the fluctuations of the brane are
unbounded implying that the brane is ripped by the strong gravity
around  the singularity. In the null case, i.e. when the
singularity is at conformal infinity, the fluctuations oscillate
in a bounded manner while converging to the singularity.
The bouncing case is equivalent to the behaviour of a brane in a
supergravity background. As we only consider intrinsic
fluctuations of the brane in an unperturbed bulk, this
corresponds to a situation where supersymmetry is preserved by
the bulk while broken by the brane motion.
Therefore the bouncing brane fluctuations correspond to
fluctuations of a non-BPS brane embedded in a supergravity background.

In the ever-expanding scenario, we can distinguish two possibilities.
The brane can escape to infinity with a scale factor which is either
expanding in a decelerating manner or accelerating, i.e.
corresponding to an inflationary era of the power law type.
In the decelerating case, the brane eventually oscillates forever.
In the inflationary case, the brane is such that any fluctuation
of a giving length scales oscillates until it freezes in while
passing through the horizon.
Of course this scenario is reminiscent of  four-dimensional
inflation modelled with a scalar field. Here the features of
inflation, i.e. the relationship between the power spectrum and
the scale factor, differ from the four dimensional case.
This is an interesting observation as it leads to a new twist in the
building of inflationary models.
One might hope that alternative scenarios to four-dimensional inflation
may emerge from five dimensional brane models and their fluctuations.


\begin{thebibliography}{99}

\bibitem{l02}
D.~Langlois,
arXiv:hep-th/0209261.

\bibitem{low}
A.~Lukas, B.~A.~Ovrut, K.~S.~Stelle and D.~Waldram
Phys.\ Rev.\ D {\bf 59}, 086001 (1999)
[arXiv:hep-th/9803235];
Nucl.\ Phys.\ B {\bf 552}, 246 (1999)
[arXiv:hep-th/9806051];
A.~Lukas, B.~A.~Ovrut and D.~Waldram,
Phys.\ Rev.\ D {\bf 60}, 086001 (1999)
[arXiv:hep-th/9806022]; 
Phys.\ Rev.\ D {\bf 61}, 023506 (2000)
[arXiv:hep-th/9902071].

\bibitem{cr99}
H.~A.~Chamblin and H.~S.~Reall,
Nucl.\ Phys.\ B {\bf 562}, 133 (1999)
[arXiv:hep-th/9903225].

\bibitem{mw00}
K.~i.~Maeda and D.~Wands,
Phys.\ Rev.\ D {\bf 62}, 124009 (2000)
[arXiv:hep-th/0008188].

\bibitem{mb00}
A.~Mennim and R.~A.~Battye,
Class.\ Quant.\ Grav.\  {\bf 18}, 2171 (2001)
[arXiv:hep-th/0008192].

\bibitem{bd}
P.~Brax and A.~C.~Davis,
Phys.\ Lett.\ B {\bf 497}, 289 (2001)
[arXiv:hep-th/0011045];
P.~Brax, C.~van de Bruck and A.~C.~Davis,
JHEP {\bf 0110}, 026 (2001)
[arXiv:hep-th/0108215].

\bibitem{fkv01}
A.~Feinstein, K.~E.~Kunze and M.~A.~V\'azquez-Mozo,
Phys.\ Rev.\ D {\bf 64}, 084015 (2001)
[arXiv:hep-th/0105182];
K.~E.~Kunze and M.~A.~V\'azquez-Mozo,
Phys.\ Rev.\ D {\bf 65}, 044002 (2002)
[arXiv:hep-th/0109038].

\bibitem{ dew} O. De Wolfe, D. Z. Freedman, S. Gubser and A. Karch Phys. \
Rev. \ D {\bf 62} (2000) 0460008 [arXiv: hep-th/9909134]

\bibitem{csa} C. Csaki, J. Erlich, C. Grojean and T. Hollowood Nucl.\ Phys. B
{\bf 584}  (2000) 359 [arXiv: hep-th/ 0004133]

\bibitem{lidsey01}
J.~E.~Lidsey,
Phys.\ Rev.\ D {\bf 64}, 063507 (2001)
[arXiv:hep-th/0106081].

\bibitem{gqtz01}
C.~Grojean, F.~Quevedo, G.~Tasinato and I.~Zavala C.,
JHEP {\bf 0108}, 005 (2001)
[arXiv:hep-th/0106120].

\bibitem{lr01}
D.~Langlois and M.~Rodr\'{\i}guez-Mart\'{\i}nez,
Phys.\ Rev.\ D {\bf 64}, 123507 (2001)
[arXiv:hep-th/0106245].

\bibitem{davis01}
S.~C.~Davis,
JHEP {\bf 0203}, 054 (2002)
[arXiv:hep-th/0106271];
JHEP {\bf 0203}, 058 (2002)
[arXiv:hep-ph/0111351].


\bibitem{charmousis01}
C.~Charmousis,
Class.\ Quant.\ Grav.\  {\bf 19}, 83 (2002)
[arXiv:hep-th/0107126].

\bibitem{ftw01}
E.~E.~Flanagan, S.~H.~Tye and I.~Wasserman,
Phys.\ Lett.\ B {\bf 522}, 155 (2001)
[arXiv:hep-th/0110070].


\bibitem{gold} W. D. Goldberger and M. B. Wise, Phys.\ Rev. Lett. {\bf 83} (1999)
4922 [arXiv: hep-th/9907447], Phys. Rev. D {\bf 60} (1999) 107505 [arXiv: hep-th 9907218]

\bibitem{gar} J. Garriga and A. Vilenkin, Phys. Rev. D {\bf 44} (1991) 1007

\bibitem{guv} J. Guven, Phys. Rev. D{\bf 48} (1993) 4604 

\bibitem{ishi} A. Ishibashi and T. Tanaka, [arXiv: gr-qc/ 0208006]

\bibitem{steer}
T. Boehm and D. A. Steer, Phys. Rev. D {\bf 66} (2002) 063510
[arXiv:hep-th/0206147].

\bibitem{static}
R.~G.~Cai, J.~Y.~Ji and K.~S.~Soh,
Phys.\ Rev.\ D {\bf 57}, 6547 (1998)
[arXiv:gr-qc/9708063].

\bibitem{ddk00}
N.~Deruelle, T.~Dolezel and J.~Katz,
Phys.\ Rev.\ D {\bf 63}, 083513 (2001)
[arXiv:hep-th/0010215].

\end{thebibliography}
\end{document}